\documentclass[12pt]{article}

\usepackage[margin=1in,nohead,dvips,a4paper]{geometry}

\usepackage{times,pstricks,pst-node}
\usepackage{amssymb,amsmath}
\usepackage[margin=1em,font=small,justification=centerlast]{caption}

\psset{linewidth=.5pt,dash=4pt 4pt,unit=.15in,arrowsize=2pt 3}

\def\arrowLine(#1,#2)(#3,#4){%
  \pcline(#1,#2)(#3,#4)%
  \lput{:U}{
    \pspicture(0,0)(0,0)
      \psline[arrows=->](2.3pt,0)(2.4pt,0)
    \endpspicture
  }
}

\def\vertexA #1{%
  \pspicture(0,0)(2,2)
    \arrowLine(0,1)(1,1)
    \arrowLine(1,1)(1,2)
    \arrowLine(2,1)(1,1)
    \arrowLine(1,1)(1,0)
    \psdot(1,1)
    \rput(0.6,0.6){#1}
  \endpspicture
}

\def\vertexB #1{%
  \pspicture(0,0)(2,2)
    \arrowLine(1,1)(0,1)
    \arrowLine(1,2)(1,1)
    \arrowLine(1,1)(2,1)
    \arrowLine(1,0)(1,1)
    \psdot(1,1)
    \rput(0.6,0.6){#1}
  \endpspicture
}

\def\vertexC #1{%
  \pspicture(0,0)(2,2)
    \arrowLine(0,1)(1,1)
    \arrowLine(1,1)(1,2)
    \arrowLine(1,1)(2,1)
    \arrowLine(1,0)(1,1)
    \psdot(1,1)
    \rput(0.6,0.6){#1}
  \endpspicture
}

\def\vertexD #1{%
  \pspicture(0,0)(2,2)
    \arrowLine(1,1)(0,1)
    \arrowLine(1,2)(1,1)
    \arrowLine(2,1)(1,1)
    \arrowLine(1,1)(1,0)
    \psdot(1,1)
    \rput(0.6,0.6){#1}
  \endpspicture
}

\def\vertexE #1{%
  \pspicture(0,0)(2,2)
    \arrowLine(1,1)(0,1)
    \arrowLine(1,1)(1,2)
    \arrowLine(2,1)(1,1)
    \arrowLine(1,0)(1,1)
    \psdot(1,1)
    \rput(0.6,0.6){#1}
  \endpspicture
}

\def\vertexF #1{%
  \pspicture(0,0)(2,2)
    \arrowLine(0,1)(1,1)
    \arrowLine(1,2)(1,1)
    \arrowLine(1,1)(2,1)
    \arrowLine(1,1)(1,0)
    \psdot(1,1)
    \rput(0.6,0.6){#1}
  \endpspicture
}

\def \Pf{\mathop{\mathrm{Pf}}}
\def\bm #1{\mbox{\boldmath $#1$}}
\def\mathsmall #1{\mbox{\small $#1$}}
\def\rmi {\mathrm i}

\begin{document}

\title{Enumeration of quarter-turn symmetric alternating-sign matrices of
odd order}
\author{A.~V.~Razumov, Yu.~G.~Stroganov\\
\small \it Institute for High Energy Physics\\[-.5em]
\small \it 142281 Protvino, Moscow region, Russia}
\date{}

\maketitle

\begin{abstract}
It was shown by Kuperberg that the partition function of the square-ice
model related to the quarter-turn symmetric alternating-sign matrices of
even order is the product of two similar factors. We propose a square-ice
model whose states are in bijection with the quarter-turn symmetric
alternating-sign matrices of odd order, and show that the partition
function of this model can be also written in a similar way. This allows to
prove, in particular, the conjectures by Robbins related to the enumeration
of the quarter-turn symmetric alternating-sign matrices.
\end{abstract}

\section{Introduction}

An alternating-sign matrix is a matrix with entries $1$, $0$, and $-1$ such
that the $1$ and $-1$ entries alternate in each column and each row and
such that the first and last nonzero entries in each row and column are
$1$. Starting  from the famous conjectures by Mills, Robbins and Rumsey
\cite{MilRobRum82, MilRobRum83} a lot of enumeration and equinumeration
results on alternating-sign matrices and their various subclasses were
obtained. Most of the results were proved using bijections between
alternating-sign matrices
and states of different variants of the statistical square-ice model. For
the first time such a method to solve enumeration problems was used by
Kuperberg~\cite{Kup96}, see also the rich in results paper~\cite{Kup02}.

Our previous paper \cite{RazStr05} is devoted to enumerations of the
half-turn symmetric alternating-sign matrices of odd order on the base of
the corresponding square-ice model. In the present paper we again treat
matrices of odd order. But this time we consider the quarter-turn
symmetric alter\-na\-ting-sign matrices.

In Section 2 we discuss first the square-ice model related to the
quarter-turn symmetric alter\-na\-ting-sign matrices of even order proposed
by Kuperberg \cite{Kup02}. Then a square-ice model whose states are in
bijection with the quarter-turn symmetric alternating-sign matrices of odd
order is introduced. In contrast with the case of the matrices of even
order, the usual recursive relations are not enough to determine the
partition function of the model recursively by Lagrange interpolation.

In Section 3 we obtain some important additional recursive relations
involving the special spectral parameter that is attached to the middle
line of the graph describing the states of the model.

In Section 4 we show that the partition function of the model is the
product of two factors closely related to the Pfaffians used by Kuperberg
to write an expression for the partition function of the square-ice model
corresponding to quarter-turn symmetric alternating-sign matrices of even
order \cite{Kup02}.

In Section 5 we consider an important special case of the overall
parameter of the model that allow to prove, in particular, the
enumeration conjectures by Robbibs \cite{Rob00} on the  quarter-turn
symmetric
alternating-sign matrices of odd order.

We denote $\bar x = x^{-1}$ and use the following convenient abbreviations
\begin{gather*}
\sigma(x) = x - \bar x, \\
\alpha(x) = \sigma(a x)\sigma(a \bar x),
\end{gather*}
proposed by Kuperberg \cite{Kup02}. Here $a$ is some parameter, which will
be introduced below.

\section{Square-ice models related to quarter-turn symmetric
al\-ter\-nating-sign matrices}

An alternating-sign $n \times n$ matrix $A$ is said to be quarter-turn
symmetric if
\[
(A)_{j,n+1-i} = (A)_{ij}, \qquad i,j = 1, \ldots n.
\]
It can be shown that
quarter-turn symmetric alternating-sign matrices of an even order $n$ exist
only when $n$ is a multiple of $4$. A quarter-turn symmetric
alternating-sign matrix of order $n = 2m+1$ has $-1$ in the center if $m$
is odd, and it has $1$ in the center if $m$ is even.

To enumerate a symmetry class of the alternating-sign matrices Kuperberg
proposed to start with a square-ice model whose states are in bijection
with the elements of the symmetry class under consideration \cite{Kup02}.
The
next step is to find the partition function of the model, defined as
the sum of the weights of all possible states. It appears that for many
symmetry classes of alternating-sign matrices a determinant or Pfaffian
representation of the partition function of the corresponding square-ice
model can be found. Using such a representation and specifying in an
appropriate way the parameters of the model one finds desired enumerations
\cite{Kup02,Zei96b,Kup96,RazStr04,RazStr05}.

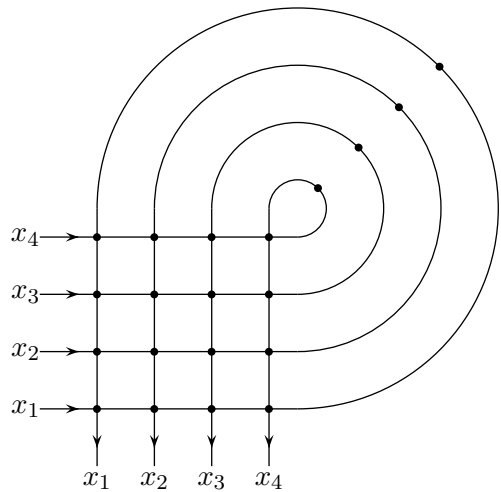
\begin{figure}[ht]
  \centering
  \begin{pspicture}(-1,-1)(16,16)
    \arrowLine(2,2)(2,0)
    \arrowLine(4,2)(4,0)
    \arrowLine(6,2)(6,0)
    \arrowLine(8,2)(8,0)
    \arrowLine(0,2)(2,2)
    \arrowLine(0,4)(2,4)
    \arrowLine(0,6)(2,6)
    \arrowLine(0,8)(2,8)
    \psline(2,2)(9,2)
    \psline(2,4)(9,4)
    \psline(2,6)(9,6)
    \psline(2,8)(9,8)
    \psline(2,2)(2,9)
    \psline(4,2)(4,9)
    \psline(6,2)(6,9)
    \psline(8,2)(8,9)
    \psarc(9,9){1}{270}{180}
    \psarc(9,9){3}{270}{180}
    \psarc(9,9){5}{270}{180}
    \psarc(9,9){7}{270}{180}
    \psdots(2,2)(2,4)(2,6)(2,8)(4,2)(4,4)(4,6)(4,8)
    \psdots(6,2)(6,4)(6,6)(6,8)(8,2)(8,4)(8,6)(8,8)
    \rput(-0.5,8){\small $x_4$}
    \rput(-0.5,6){\small $x_3$}
    \rput(-0.5,4){\small $x_2$}
    \rput(-0.5,2){\small $x_1$}
    \rput(2,-0.5){\small $x_1$}
    \rput(4,-0.5){\small $x_2$}
    \rput(6,-0.5){\small $x_3$}
    \rput(8,-0.5){\small $x_4$}
    \SpecialCoor
    \psdots[origin={-9,-9}](1;45)(3;45)(5;45)(7;45)
\end{pspicture}
\caption{Square-ice corresponding to the quarter-turn symmetric
alternating-sign matrices of even order}
\label{f:1}
\end{figure}

To describe the states of a square-ice model it is convenient to use a
graphical pattern. For example, the states of the square-ice model
corresponding to the quarter-turn symmetric alternating-sign matrices of an
even order are described by the graph given in Figure \ref{f:1}.
The labels $x_i$ are the spectral parameters which are used
to define the partition function of the model. To get a concrete state of
the model one chooses an orientation for each of the unoriented edges in
such a way that two edges enter and leave every tetravalent vertex, and
either two edges enter or two edges leave every bivalent
vertex.\footnote{Kuperberg uses a dashed line crossing an edge to say
that its orientaion reverses as it crosses the line \cite{Kup02}. It is
convenient for our purposes to treat reversal of the orientation as a
special type of a vertex.} Certainly, we draw a pattern for a fixed order
of matrices, but a generalisation to the case of an arbitrary possible
order is always evident.

\begin{figure}[ht]
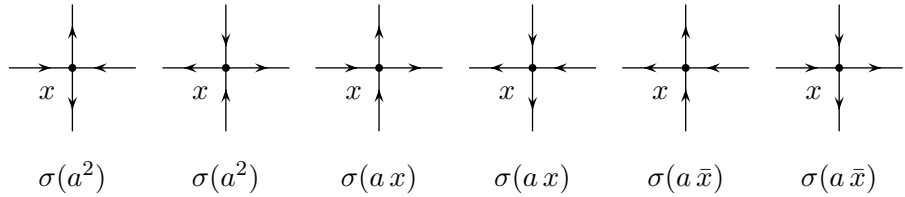

\[
\psset{unit=2em}
\begin{array}{cccccc}
\vertexA{\small $x$} & \vertexB{\small $x$} & \vertexC{\small $x$} &
\vertexD{\small $x$} & \vertexE{\small $x$} & \vertexF{\small $x$} \\[.5em]
\mathsmall{\sigma(a^2)} & \mathsmall{\sigma(a^2)} & \mathsmall{\sigma(a \,
x)} & \mathsmall{\sigma(a \, x)} & \mathsmall{\sigma(a \,\bar x)} &
\mathsmall{\sigma(a \, \bar x)}
\end{array}
\]
\caption{The weights of the tetravalent vertices}
\label{f:wghts}
\end{figure}

The weight of a state is the product of the weights of the vertices. The
choice for the weights of tetravalent vertices used in the present paper is
as given in Figure \ref{f:wghts}.
The parameter $a$ is common for all tetravalent vertices. All bivalent
vertices have weight $1$. If a vertex is unlabelled and formed by
intersection of two labeled lines, then the value of the vertex label is
set to $x \bar y$ if it is in the quadrant which is swept by the line with
the spectral parameter $x$ when it is rotated anticlockwise to the line
with the spectral parameter $y$. One may move a vertex label $x$ one
quadrant to an adjacent one changing it to $\bar x$.

A graph, similar to the one given in Figure \ref{f:1}, denotes also the
corresponding function. Here the summation over all possible orientations
of internal edges is implied. It can be easily understood that our
conventions make the formalism invariant under rotations and orientation
preserving smooth deformation of graphs. If we reflect a graph over a line
and overline the line labels we obtain a graph which describes the same
function as the initial graph. After all, reversing orientation of all
oriented edges we obtain a graph which again gives the same function as the
initial graph.

If we have unoriented boundary edges, then the graph represents the set of
the quantities corresponding to their possible orientations. Usually,
graphs with such edges arise when we give a graphical representation of
equality of functions. In such a case, if it is needed, we should rotate
both sides of an equality simultaneously.

As a useful example one can take the graph
corresponding to the well-known Yang--Baxter equation
\begin{equation}
\psset{unit=.25in}
\begin{pspicture}[.45](-.5,-.5)(4.5,4.5)
  \pscurve(0,1)(1.3,2.0)(2.6,2.6)(4,3)
  \pscurve(0,3)(1.3,2.0)(2.6,1.4)(4,1)
  \pscurve(2.5,4)(2.8,2.7)(2.8,1.3)(2.5,0)
  \rput(.5,2){\small $z$}
  \rput(3.3,3.3){\small $y$}
  \rput(3.3,.7){\small $x$}
  \psdots(1.3,2)(2.8,1.34)(2.8,2.66)
\end{pspicture}
\quad = \quad
\begin{pspicture}[.45](-.5,-.5)(4.5,4.5)
  \pscurve(4,1)(2.7,2.0)(1.4,2.6)(0,3)
  \pscurve(4,3)(2.7,2.0)(1.4,1.4)(0,1)
  \pscurve(1.5,4)(1.2,2.7)(1.2,1.3)(1.5,0)
  \rput(3.5,2){\small $z$}
  \rput(.7,3.3){\small $x$}
  \rput(.7,.7){\small $y$}
  \psdots(2.7,2)(1.2,1.34)(1.2,2.66)
\end{pspicture} \label{e:yb}
\end{equation}
This equation is satisfied if $xyz = a$.

The procedure described above to find enumerations of  quarter-turn
symmetric
alternating-sign matrices of even order was realised by Kuperberg
\cite{Kup02}. In the present paper we treat the case of  quarter-turn
symmetric alternating-sign matrices of odd order. The graphical pattern for
the state space of the corresponding square-ice model depends on the order
of the matrices. For the order $2m+1$ with $m$ even we have the pattern
given at Figure \ref{f:3}, and for the order $2m+1$ with $m$ odd we have
the pattern given in Figure \ref{f:4}.
\begin{figure}[ht]
  \centering
  \begin{minipage}[t]{.425\linewidth}
    \centering
    \begin{pspicture}(-1,-1)(10,10)
      \arrowLine(2,2)(2,0)
      \arrowLine(4,2)(4,0)
      \arrowLine(6,6)(6,4)
      \arrowLine(0,2)(2,2)
      \arrowLine(0,4)(2,4)
      \arrowLine(6,2)(6,0)
      \psline(2,2)(6,2)
      \psline(2,4)(6,4)
      \psline(2,2)(2,6)
      \psline(4,2)(4,6)
      \psline(6,4)(6,2)
      \psdots(2,2)(2,4)(4,2)(4,4)(6,4)(6,2)
      \rput(-0.5,4){\small $x_2$}
      \rput(-0.5,2){\small $x_1$}
      \rput(2,-0.5){\small $x_1$}
      \rput(4,-0.5){\small $x_2$}
      \rput(6,-0.5){\small $x_3$}
      \psarc(6,6){2}{270}{180}
      \psarc(6,6){4}{270}{180}
      \SpecialCoor
      \psdots[origin={-6,-6}](2;45)(4;45)
    \end{pspicture}
    \caption{Square-ice corresponding to quarter-turn symmetric
             alternating-sign matrices of order 5} \label{f:3}
  \end{minipage}
  \hspace{.05\linewidth}
  \begin{minipage}[t]{.425\linewidth}
    \centering
    \begin{pspicture}(-1,-1)(14,14)
      \arrowLine(2,2)(2,0)
      \arrowLine(4,2)(4,0)
      \arrowLine(6,2)(6,0)
      \arrowLine(8,2)(8,0)
      \arrowLine(8,6)(8,8)
      \arrowLine(0,2)(2,2)
      \arrowLine(0,4)(2,4)
      \arrowLine(0,6)(2,6)
      \psline(2,2)(8,2)
      \psline(2,4)(8,4)
      \psline(2,6)(8,6)
      \psline(2,2)(2,8)
      \psline(4,2)(4,8)
      \psline(6,2)(6,8)
      \psline(8,6)(8,2)
      \psdots(2,2)(2,4)(2,6)(4,2)(4,4)(4,6)
      \psdots(6,2)(6,4)(6,6)(8,2)(8,4)(8,6)
      \rput(-0.5,6){\small $x_3$}
      \rput(-0.5,4){\small $x_2$}
      \rput(-0.5,2){\small $x_1$}
      \rput(2,-0.5){\small $x_1$}
      \rput(4,-0.5){\small $x_2$}
      \rput(6,-0.5){\small $x_3$}
      \rput(8,-0.5){\small $x_4$}
      \psarc(8,8){2}{270}{180}
      \psarc(8,8){4}{270}{180}
      \psarc(8,8){6}{270}{180}
      \SpecialCoor
      \psdots[origin={-8,-8}](2;45)(4;45)(6;45)
    \end{pspicture}
    \caption{Square-ice corresponding to the quarter-turn symmetric
             alternating-sign matrices of order 7} \label{f:4}
  \end{minipage}
\end{figure}
The difference actually is in the orientation of the boundary edges
belonging to the `middle' line. It is not difficult to get convinced that
there is a bijection between the states of the square-ice models described
by Figures \ref{f:3} and \ref{f:4} and the corresponding subsets of the
alternating-sign matrices.

The partition function of the model depends on $m+1$ spectral parameters
$x_1, x_2, \ldots, x_{m+1}$. We denote it $Z_{\mathrm{QT}}(2m + 1; \bm x)$,
where $\bm x = (x_1, \ldots, x_{m+1})$ is the $(m+1)$-dimensional vector
formed by the spectral parameters. Using Yang--Baxter equation (\ref{e:yb})
and an evident equality
\[
\begin{pspicture}[.4](-1,-1)(4,3)
  \psbezier(0,2)(1,2)(2,0)(3,0)
  \psbezier(0,0)(1,0)(2,2)(3,2)
  \psline(3,0)(4,0)
  \psline(3,2)(4,2)
  \psdots(1.5,1)(3,0)(3,2)
  \rput(-.5,0){\small $x$}
  \rput(-.5,2){\small $y$}
\end{pspicture}
\quad = \quad
\begin{pspicture}[.4](-1,-1)(4,3)
  \psbezier(1,2)(2,2)(3,0)(4,0)
  \psbezier(1,0)(2,0)(3,2)(4,2)
  \psline(0,0)(1,0)
  \psline(0,2)(1,2)
  \psdots(2.5,1)(1,0)(1,2)
  \rput(-.5,0){\small $x$}
  \rput(-.5,2){\small $y$}
\end{pspicture} \quad
\]
one can show that the function $Z_{\mathrm{QT}}(2m + 1; \bm x)$ is
symmetric in the variables $x_1, x_2, \ldots, x_m$.

It is not difficult to get convinced that the following equality
\begin{equation}
\begin{pspicture}[.5](-1,-2)(6,7)
  \psline(0,2)(6,2)
  \psline(0,4)(6,4)
  \psline(2,2)(2,4)
  \arrowLine(2,2)(2,0)
  \arrowLine(2,4)(2,6)
  \psdots(2,2)(2,4)(4,2)(4,4)
  \rput(-.5,2){\small $x$}
  \rput(-.5,4){\small $y$}
  \rput(2,-.5){\small $z$}
\end{pspicture}
\quad = \quad
\begin{pspicture}[.5](-1,-2)(6,7)
  \psline(0,2)(6,2)
  \psline(0,4)(6,4)
  \psline(4,2)(4,4)
  \arrowLine(4,0)(4,2)
  \arrowLine(4,6)(4,4)
  \psdots(2,2)(2,4)(4,2)(4,4)
  \rput(-.5,2){\small $x$}
  \rput(-.5,4){\small $y$}
  \rput(4,-.5){\small $z$}
\end{pspicture}
\label{e:2}
\end{equation}
is valid. Actually there are similar equalities with different orientations
of the oriented edges in the left hand side and reversed orientations of
the corresponding edges in the right hand side. Reflect now the graph in
Figure \ref{f:4} over the line which is drawn as a dotted line in Figure
\ref{f:5},
\begin{figure}[ht]
  \centering
  \begin{minipage}[t]{.425\linewidth}
    \centering
    \begin{pspicture}(-1,-1)(14,14)
      \psline[linestyle=dotted](1,1)(13,13)
      \arrowLine(2,2)(2,0)
      \arrowLine(4,2)(4,0)
      \arrowLine(6,2)(6,0)
      \arrowLine(0,2)(2,2)
      \arrowLine(0,4)(2,4)
      \arrowLine(0,6)(2,6)
      \arrowLine(0,8)(2,8)
      \arrowLine(8,8)(6,8)
      \psline(2,2)(8,2)
      \psline(2,4)(8,4)
      \psline(2,6)(8,6)
      \psline(2,8)(6,8)
      \psline(2,2)(2,8)
      \psline(4,2)(4,8)
      \psline(6,2)(6,8)
      \psdots(2,2)(2,4)(2,6)(4,2)(4,4)(4,6)
      \psdots(6,2)(6,4)(6,6)(2,8)(4,8)(6,8)
      \rput(-0.5,6){\small $\bar x_3$}
      \rput(-0.5,4){\small $\bar x_2$}
      \rput(-0.5,2){\small $\bar x_1$}
      \rput(2,-0.5){\small $\bar x_1$}
      \rput(4,-0.5){\small $\bar x_2$}
      \rput(6,-0.5){\small $\bar x_3$}
      \rput(-.5,8){\small $\bar x_4$}
      \psarc(8,8){2}{270}{180}
      \psarc(8,8){4}{270}{180}
      \psarc(8,8){6}{270}{180}
      \SpecialCoor
      \psdots[origin={-8,-8}](2;45)(4;45)(6;45)
    \end{pspicture}
    \caption{} \label{f:5}
  \end{minipage}
  \hspace{.05\linewidth}
  \begin{minipage}[t]{.425\linewidth}
    \centering
    \begin{pspicture}(-1,-1)(14,14)
      \arrowLine(2,2)(2,0)
      \arrowLine(4,2)(4,0)
      \arrowLine(6,2)(6,0)
      \arrowLine(8,2)(8,0)
      \arrowLine(8,6)(8,8)
      \arrowLine(0,2)(2,2)
      \arrowLine(0,4)(2,4)
      \arrowLine(0,6)(2,6)
      \psline(2,2)(8,2)
      \psline(2,4)(8,4)
      \psline(2,6)(8,6)
      \psline(2,2)(2,8)
      \psline(4,2)(4,8)
      \psline(6,2)(6,8)
      \psline(8,6)(8,2)
      \psdots(2,2)(2,4)(2,6)(4,2)(4,4)(4,6)
      \psdots(6,2)(6,4)(6,6)(8,2)(8,4)(8,6)
      \rput(-0.5,6){\small $\bar x_3$}
      \rput(-0.5,4){\small $\bar x_2$}
      \rput(-0.5,2){\small $\bar x_1$}
      \rput(2,-0.5){\small $\bar x_1$}
      \rput(4,-0.5){\small $\bar x_2$}
      \rput(6,-0.5){\small $\bar x_3$}
      \rput(8,-0.5){\small $\bar x_4$}
      \psarc(8,8){2}{270}{180}
      \psarc(8,8){4}{270}{180}
      \psarc(8,8){6}{270}{180}
      \SpecialCoor
      \psdots[origin={-8,-8}](2;45)(4;45)(6;45)
    \end{pspicture}
    \caption{} \label{f:6}
  \end{minipage}
\end{figure}
overline all the labels, and reverse the orientations of the oriented
edges. As follows from the remarks made above, the resulting graph which
is given in Figure \ref{f:5} corresponds to the same function as the
graph given in Figure \ref{f:4}. Using equality (\ref{e:2}), we transform
Figure \ref{f:5} to Figure \ref{f:6} which again corresponds to the same
function as the graph given in Figure \ref{f:4}. Thus, we proved the
equality
\begin{equation}
Z_{\mathrm{QT}}(2m + 1; \bm x) = Z_{\mathrm{QT}}(2m + 1; \bar{\bm x}),
\label{e:3}
\end{equation}
where $\bar{\bm x} = (\bar x_1, \ldots, \bar x_{m+1})$.

Following the usual procedure (see, for example, the proof of Lemma 13 in
paper \cite{Kup02}) we obtain $2m - 2$ recursive relations
\begin{align*}
Z_{\mathrm{QT}}(2m + 1; \bm x)|_{x_1 = a x_j} = \sigma^{2}(a) &
\sigma^{2}(a^2) \\
& \times \prod_{\substack{k = 2 \\ k \ne j}}^{m+1} \sigma^2(a^2 \bar
x_k x_j) \sigma^2(a \bar x_j x_k)  Z_{\mathrm{QT}}(2m - 3; \bm x
\smallsetminus
x_1 \smallsetminus x_j), \\
Z_{\mathrm{QT}}(2m + 1; \bm x)|_{x_1 = \bar a x_j} = \sigma^{2}(a) &
\sigma^{2}(a^2) \\
& \times \prod_{\substack{k = 2 \\ k \ne j}}^{m+1} \sigma^2(a \bar x_k x_j)
\sigma^2(a^2 \bar x_j x_k)  Z_{\mathrm{QT}}(2m - 3; \bm x \smallsetminus
x_1 \smallsetminus x_j),
\end{align*}
where $j = 2, \ldots, m$. Since the partition function $Z_{\mathrm{QT}}(2m
+ 1; \bm x)$ is symmetric in the variables $x_1, \ldots, x_m$, we actually
have $m^2 - m$ recursive relations
\begin{multline}
Z_{\mathrm{QT}}(2m + 1; \bm x)|_{x_i = a x_j} =
\sigma^{2}(a) \sigma^{2}(a^2) \\
\times \prod_{\substack{k=1 \\ k \ne i,j}}^{m+1} \sigma^2(a^2 \bar x_k x_j)
\sigma^2(a \bar x_j x_k)  Z_{\mathrm{QT}}(2m - 3; \bm x \smallsetminus
x_i \smallsetminus x_j),
\label{1st}
\end{multline}
where $i, j=1, \ldots, m$, and $i \ne j$.

Consider some fixed value of the index $i$ such that $1 \le i \le m$. The
partition function $Z_{\mathrm{QT}}(2m + 1; \bm x)$ is a centered Laurent
polynomial of width $2m - 2$ in the square of the variable $x_i$.
Therefore, if we know $2m-1$ values of $Z_{\mathrm{QT}}(2m + 1; \bm x)$ for
$2m-1$ values of $x_i^2$ we know $Z_{\mathrm{QT}}(2m + 1; \bm x)$
completely. Recursive relations (\ref{1st}) supply us with the expressions
for $Z_{\mathrm{QT}}(2m + 1; \bm x)$ via $Z_{\mathrm{QT}}(2m - 3; \bm x
\smallsetminus x_i \smallsetminus x_j)$ for $2m - 2$ values of $x_i$. It is
not enough to determine $Z_{\mathrm{QT}}(2m + 1; \bm x)$ recursively by the
Lagrange interpolation.\footnote{Note that for $Z_{\mathrm{QT}}(4k, \bm x)$
the recursive relations similar to (\ref{1st}) give enough data for
Lagrange interpolation \cite{Kup02}.} From the other hand, the partition
function
$Z_{\mathrm{QT}}(2m + 1; \bm x)$ is a centered Laurent polynomial in
$x_{m+1}^2$ of width $m-1$ if $m$ is odd, and of width $m$ if $m$ is even.
It appears that there are enough recursive relations involving the variable
$x_{m+1}$ to determine $Z_{\mathrm{QT}}(2m + 1; \bm x)$ via Lagrange
interpolation.

\section{\mathversion{bold}Recursive relations involving variable
$x_{m+1}$}

Multiply the partition function $Z_{\mathrm{QT}}(2m + 1; \bm x)$ by
$\sigma(a z_m)$, where $z_m$ is a parameter which is not specified yet. One
can easily understood that the resulting function can be represented by
Figure~\ref{f:7}.
\begin{figure}[ht]
  \centering
  \begin{minipage}[t]{.425\linewidth}
    \centering
    \begin{pspicture}(-3,-3)(14,14)
      \arrowLine(2,2)(2,0)
      \arrowLine(4,2)(4,0)
      \arrowLine(8,6)(8,8)
      \arrowLine(0,2)(2,2)
      \arrowLine(0,4)(2,4)
      \arrowLine(0,6)(2,6)
      \psline(2,2)(8,2)
      \psline(2,4)(8,4)
      \psline(2,6)(8,6)
      \psline(2,2)(2,8)
      \psline(4,2)(4,8)
      \psline(6,2)(6,8)
      \psline(8,6)(8,2)
      \psdots(2,2)(2,4)(2,6)(4,2)(4,4)(4,6)
      \psdots(6,2)(6,4)(6,6)(8,2)(8,4)(8,6)
      \rput(-0.5,6){\small $x_3$}
      \rput(-0.5,4){\small $x_2$}
      \rput(-0.5,2){\small $x_1$}
      \rput(2,-0.5){\small $x_1$}
      \rput(4,-0.5){\small $x_2$}
      \rput(8,-2.5){\small $x_3$}
      \rput(6,-2.5){\small $x_4$}
      \rput(7,-.5){\small $z_3$}
      \psarc(8,8){2}{270}{180}
      \psarc(8,8){4}{270}{180}
      \psarc(8,8){6}{270}{180}
      \psbezier(8,2)(8,1)(6,0)(6,-1)
      \psbezier(6,2)(6,1)(8,0)(8,-1)
      \psline[arrows=<](8,-1)(8,-2)
      \psline(8,-1)(8,-1.3)
      \psline[arrows=<](6,-1)(6,-2)
      \psline(6,-1)(6,-1.3)
      \psdot(7,.5)
      \SpecialCoor
      \psdots[origin={-8,-8}](2;45)(4;45)(6;45)
    \end{pspicture}
    \caption{} \label{f:7}
  \end{minipage}
  \hspace{.05\linewidth}
  \begin{minipage}[t]{.425\linewidth}
    \centering
    \begin{pspicture}(-1,-1)(16,16)
      \arrowLine(2,2)(2,0)
      \arrowLine(4,2)(4,0)
      \arrowLine(6,2)(6,0)
      \arrowLine(8,2)(8,0)
      \arrowLine(8,8)(10,8)
      \arrowLine(0,2)(2,2)
      \arrowLine(0,4)(2,4)
      \arrowLine(0,6)(2,6)
      \psline(2,2)(9,2)
      \psline(2,4)(9,4)
      \psline(2,6)(10,6)
      \psline(2,2)(2,9)
      \psline(4,2)(4,9)
      \psline(6,2)(6,7)
      \psline(8,8)(8,2)
      \psline(7,8)(8,8)
      \psdots(2,2)(2,4)(2,6)(4,2)(4,4)(4,6)
      \psdots(6,2)(6,4)(6,6)(8,2)(8,4)(8,6)(8,8)
      \rput(-0.5,6){\small $x_3$}
      \rput(-0.5,4){\small $x_2$}
      \rput(-0.5,2){\small $x_1$}
      \rput(2,-0.5){\small $x_1$}
      \rput(4,-0.5){\small $x_2$}
      \rput(6,-0.5){\small $x_4$}
      \rput(8,-0.5){\small $x_3$}
      \rput(7.4,7.4){\small $z_3$}
      \psarc(7,7){1}{90}{180}
      \psarc(10,8){2}{270}{180}
      \psarc(9,9){5}{270}{180}
      \psarc(9,9){7}{270}{180}
      \SpecialCoor
      \psdots[origin={-9,-9}](5;45)(7;45)
      \psdots[origin={-10,-8}](2;45)
    \end{pspicture}
    \caption{} \label{f:8}
  \end{minipage}
\end{figure}
Put $z_m = a x_{m+1} \bar x_m$, and transform the graph in
Figure~\ref{f:7} to the graph in Figure~\ref{f:8} using the
Yang--Baxter equation~(\ref{e:yb}). Repeating this procedure we see that
Figure~\ref{f:9}, where $z_i = a x_{m+1} \bar x_i$ corresponds to the
partition function $Z_{\mathrm{QT}}(2m + 1; \bm x)$ multiplied by the
product $\prod_{i=1}^m \sigma(a^2 x_{m+1} \bar x_i)$.
\begin{figure}[ht]
  \centering
  \begin{minipage}[t]{.425\linewidth}
    \centering
    \begin{pspicture}(-1,-1)(16,14)
      \arrowLine(2,2)(2,0)
      \arrowLine(4,2)(4,0)
      \arrowLine(6,2)(6,0)
      \arrowLine(8,2)(8,0)
      \arrowLine(8,8)(10,8)
      \arrowLine(0,2)(2,2)
      \arrowLine(0,4)(2,4)
      \arrowLine(0,6)(2,6)
      \psline(2,2)(10,2)
      \psline(2,4)(10,4)
      \psline(2,6)(10,6)
      \psline(2,2)(2,7)
      \psline(4,2)(4,8)
      \psline(6,2)(6,8)
      \psline(8,2)(8,8)
      \psline(3,8)(8,8)
      \psdots(2,2)(2,4)(2,6)(4,2)(4,4)(4,6)
      \psdots(6,2)(6,4)(6,6)(8,2)(8,4)(8,6)
      \psdots(8,8)(6,8)(4,8)
      \rput(-0.5,6){\small $x_3$}
      \rput(-0.5,4){\small $x_2$}
      \rput(-0.5,2){\small $x_1$}
      \rput(2,-0.5){\small $x_4$}
      \rput(4,-0.5){\small $x_1$}
      \rput(6,-0.5){\small $x_2$}
      \rput(8,-0.5){\small $x_3$}
      \rput(7.4,7.4){\small $z_3$}
      \rput(5.4,7.4){\small $z_2$}
      \rput(3.4,7.4){\small $z_1$}
      \psarc(3,7){1}{90}{180}
      \psarc(10,8){2}{270}{180}
      \psarc(10,8){4}{270}{180}
      \psarc(10,8){6}{270}{180}
      \SpecialCoor
      \psdots[origin={-10,-8}](2;45)(4;45)(6;45)
    \end{pspicture}
    \caption{} \label{f:9}
  \end{minipage}
  \hspace{.05\linewidth}
  \begin{minipage}[t]{.425\linewidth}
    \centering
    \begin{pspicture}(-1,-1)(16,14)
      \arrowLine(2,2)(2,0)
      \arrowLine(4,2)(4,0)
      \arrowLine(6,2)(6,0)
      \arrowLine(8,2)(8,0)
      \arrowLine(8,8)(10,8)
      \arrowLine(0,2)(2,2)
      \arrowLine(0,4)(2,4)
      \arrowLine(0,6)(2,6)
      \arrowLine(10,2)(8,2)
      \arrowLine(8,2)(6,2)
      \arrowLine(6,2)(4,2)
      \arrowLine(4,2)(2,2)
      \arrowLine(2,4)(4,4)
      \arrowLine(4,4)(6,4)
      \arrowLine(2,6)(4,6)
      \arrowLine(4,6)(6,6)
      \arrowLine(4,8)(6,8)
      \arrowLine(2,2)(2,4)
      \arrowLine(2,4)(2,6)
      \arrowLine(4,8)(4,6)
      \arrowLine(4,6)(4,4)
      \arrowLine(4,4)(4,2)
      \arrowLine(6,4)(6,2)
      \arrowLine(8,4)(8,2)
      \psline(6,4)(10,4)
      \psline(6,6)(10,6)
      \psline(2,6)(2,7)
      \psline(6,4)(6,8)
      \psline(8,4)(8,8)
      \psline(3,8)(4,8)
      \psline(6,8)(8,8)
      \psdots(2,2)(2,4)(2,6)(4,2)(4,4)(4,6)
      \psdots(6,2)(6,4)(6,6)(8,2)(8,4)(8,6)
      \psdots(8,8)(6,8)(4,8)
      \rput(-0.5,6){\small $x_3$}
      \rput(-0.5,4){\small $x_2$}
      \rput(-0.5,2){\small $x_1$}
      \rput(2,-0.5){\small $x_4$}
      \rput(4,-0.5){\small $x_1$}
      \rput(6,-0.5){\small $x_2$}
      \rput(8,-0.5){\small $x_3$}
      \rput(7.4,7.4){\small $z_3$}
      \rput(5.4,7.4){\small $z_2$}
      \rput(3.4,7.4){\small $z_1$}
      \psarc(3,7){1}{90}{180}
      \psarc{<-}(3,7){1}{135}{180}
      \psarc(10,8){2}{270}{180}
      \psarc(10,8){4}{270}{180}
      \psarc{->}(10,8){6}{270}{175}
      \psarc(10,8){6}{170}{180}
      \SpecialCoor
      \psdots[origin={-10,-8}](2;45)(4;45)(6;45)
    \end{pspicture}
    \caption{} \label{f:10}
  \end{minipage}
\end{figure}
Put now $x_{m+1} = \bar a x_1$. One can easily see that after that some
vertices become fixed (see Figure \ref{f:10}). If we remove these vertices
we will come to the function described by Figure \ref{f:11}.
\begin{figure}
  \centering
  \begin{minipage}[t]{.425\linewidth}
    \centering
    \begin{pspicture}(1,1)(12,12)
      \arrowLine(6,8)(8,8)
      \arrowLine(4,4)(4,2)
      \arrowLine(6,4)(6,2)
      \arrowLine(2,6)(4,6)
      \arrowLine(2,8)(4,8)
      \arrowLine(2,4)(4,4)
      \psline(4,4)(8,4)
      \psline(2,6)(8,6)
      \psline(4,2)(4,8)
      \psline(6,2)(6,8)
      \psline(4,8)(6,8)
      \psdots(4,8)(6,8)(4,4)(4,6)(6,4)(6,6)
      \rput(1.5,6){\small $x_3$}
      \rput(1.5,4){\small $x_2$}
      \rput(1.5,8){\small $x_1$}
      \rput(4,1.5){\small $x_2$}
      \rput(6,1.5){\small $x_3$}
      \psarc(8,8){2}{270}{180}
      \psarc(8,8){4}{270}{180}
      \SpecialCoor
      \psdots[origin={-8,-8}](2;45)(4;45)
    \end{pspicture}
    \caption{} \label{f:11}
  \end{minipage}
  \hspace{.05\linewidth}
  \begin{minipage}[t]{.425\linewidth}
    \centering
    \begin{pspicture}(-1,-1)(10,10)
      \arrowLine(2,2)(2,0)
      \arrowLine(4,2)(4,0)
      \arrowLine(6,6)(6,4)
      \arrowLine(0,2)(2,2)
      \arrowLine(0,4)(2,4)
      \psline(2,2)(6,2)
      \psline(2,4)(6,4)
      \psline(2,2)(2,6)
      \psline(4,2)(4,6)
      \psdots(2,2)(2,4)(4,2)(4,4)(6,4)(6,2)
      \psline(6,4)(6,2)
      \arrowLine(6,2)(6,0)
      \rput(-0.5,4){\small $x_3$}
      \rput(-0.5,2){\small $x_2$}
      \rput(2,-0.5){\small $x_2$}
      \rput(4,-0.5){\small $x_3$}
      \rput(6,-0.5){\small $x_1$}
      \psarc(6,6){2}{270}{180}
      \psarc(6,6){4}{270}{180}
      \SpecialCoor
      \psdots[origin={-6,-6}](2;45)(4;45)
    \end{pspicture}
    \caption{} \label{f:12}
  \end{minipage}
\end{figure}
Removal of a fixed vertex from a graph describing a function is equivalent
to the
division of the function by the weight of the vertex. Taking into account
all multiplications and divisions we made, we obtain an important
recursive relation
\begin{multline*}
Z_{\mathrm{QT}}(2m + 1; \bm x)|_{x_{m+1} = \bar a x_1} = \sigma(a)
\sigma(a^2) \\
\times \prod_{k=2}^m \sigma(a^2 \bar x_1 x_k) \sigma(a \bar x_k x_1)
Z_{\mathrm{QT}}(2m - 1; (x_2, \ldots, x_m, x_1)).
\end{multline*}
Using the symmetricity of the partition function $Z_{\mathrm{QT}}(2m + 1;
\bm x)$ in the variables $x_1, \ldots, x_m$ we obtain $m$ recursive
relations
\begin{multline}
Z_{\mathrm{QT}}(2m + 1; \bm x)|_{x_{m+1} = \bar a x_j} = \sigma(a)
\sigma(a^2) \\
\times \prod_{\substack{k = 1 \\ k \ne j}}^m \sigma(a^2 \bar x_j x_k)
\sigma(a \bar x_k x_j) Z_{\mathrm{QT}}(2m - 1; (x_1, \ldots, \hat x_j,
\ldots, x_m, x_j)), \label{2nd}
\end{multline}
where the hat means omission of the corresponding argument. Taking into
account the inversion symmetry (\ref{e:3}), we obtain $m$ additional
recursive relations
\begin{multline}
Z_{\mathrm{QT}}(2m + 1; \bm x)|_{x_{m+1} = a x_j} = \sigma(a)
\sigma(a^2) \\
\times \prod_{\substack{k = 1 \\ k \ne j}}^m \sigma(a^2 \bar x_k x_j)
\sigma(a \bar x_j x_k) Z_{\mathrm{QT}}(2m - 1; (x_1, \ldots, \hat x_j,
\ldots, x_m, x_j)). \label{3rd}
\end{multline}
Hence we have  $2m$ specializations in the square of the variable
$x_{m+1}$. It is more than enough to reconstruct the partition function by
recursion. Certainly, we have to use also the initial value
\begin{equation}
Z_{\mathrm{QT}}(3; \bm x) = \sigma(a) \sigma(a^2). \label{ini}
\end{equation}

\section{Kuperberg's pfaffians and partition function}

Following Kuperberg for any positive integer $r$ introduce an antisymmetric
$2l \times 2l$ matrix $M^{(r)}(l; \bm x)$ with the matrix elements
\[
M_{ij}^{(r)}(l; \bm x) = \frac{\sigma(\bar x_i^r x_j^r)}{\alpha(\bar x_i
x_j)}, \qquad i, j = 1, \ldots, 2l.
\]
Recall that the Pfaffian of an antisymmetric $2l \times 2l$ matrix
$A$ can be defined as
\[
\Pf A = \frac{1}{2^l l!} \sum_{s \in S_{2l}} \mathrm{sgn}(s) \,
A_{s(1)s(2)} A_{s(3)s(4)} \ldots A_{s(2l-1)s(2l)},
\]
where $S_{2l}$ is the symmetric group of degree $2l$.

Again following Kuperberg define the following functions of $2l$ variables
$x_1,x_2,...,x_{2l}$:
\[
Z^{(r)}_{\mathrm{QT}}(l; \bm x) = \prod_{1 \leq i < j \leq 2l}
\frac{\alpha (\bar x_i x_j)} {\sigma (\bar x_i x_j)} \, \Pf M^{(r)}(l;
\bm x).
\]
The functions $Z^{(r)}_{\mathrm{QT}}(l; \bm x)$ are symmetric in the
variables $x_1, \ldots, x_{2l}$, and one can verify the validity of the
following recursive relations
\begin{equation}
\left. Z^{(r)}_{\mathrm{QT}}(l; \bm x) \right|_{x_i = a x_j} =
\frac{\sigma(a^r)}{\sigma(a)} \prod_{\substack{k = 1 \\ k \ne i, j}}^{2l}
\left[ \sigma(a^2 \bar x_k x_j) \sigma(a \bar x_j x_k) \right]
Z^{(r)}_{\mathrm{QT}}(l - 1; \bm x \smallsetminus x_i \smallsetminus x_j),
\label{e:7}
\end{equation}
where $i,j = 1, \ldots, 2l$ and $i \ne j$. On the basis of these recursive
relation Kuperberg proved \cite{Kup02} that the partition function of the
square-ice model corresponding to the quarter-turn symmetric
alternating-sign matrices of even order can be represented as
\[
Z_{\mathrm{QT}}(4l; \bm x) = [\sigma^{3l}(a) \sigma^l(a^2)]
Z^{(1)}_{\mathrm{QT}}(l; \bm x) Z^{(2)}_{\mathrm{QT}}(l; \bm x).
\]
It appears that the partition function for the case of the quarter-turn
symmetric alternating-sign matrices of odd order can be also written in a
similar way.

The function $Z^{(1)}_{\mathrm{QT}}(l; \bm x)$ is a centered Laurent
polynomial of width $2l-2$ in the square of each of the variables
$x_1, \ldots x_{2l}$, and for the function $Z^{(2)}_{\mathrm{QT}}(l; \bm
x)$ we have
\[
Z^{(2)}_{\mathrm{QT}}(l, \bm x) = \sum_{i=1}^{2l} c_i(l; \bm x
\smallsetminus x_{2l}) \, x_{2l}^{2i - 2l -1}.
\]
Introduce the function
\[
\widetilde Z^{(2)}_{\mathrm{QT}}(l, \bm x) = \left[ \prod_{k=1}^{2l-1} x_k
\right] c_{2l}(l; \bm x),
\]
which is a centered Laurent polynomial of width $2l-2$ in the square of
each the variables  $x_1, \ldots, x_{2l-1}$. It follows from (\ref{e:7})
that
\begin{equation}
\left. \widetilde Z^{(2)}_{\mathrm{QT}}(l; \bm x) \right|_{x_i = a x_j} =
- \frac{\sigma(a^2)}{\sigma(a)} \prod_{\substack{k = 1 \\ k \ne i,
j}}^{2l-1} \left[ \sigma(a^2 \bar x_k x_j) \sigma(a \bar x_j x_k) \right]
\widetilde Z^{(2)}_{\mathrm{QT}}(l - 1; \bm x \smallsetminus x_i
\smallsetminus x_j),
\label{2ndz}
\end{equation}
where $i, j = 1, \ldots, 2l-1$ and $i \ne j$. Now we can write the
following expressions the partition function of the square-ice model
corresponding to the quarter-turn symmetric alternating-sign matrices of
odd order
\begin{align}
& Z_{\mathrm{QT}}(4l + 1; \bm x) = [(-1)^l \sigma^{3l}(a) \sigma^l(a^2)]
Z^{(1)}_{\mathrm{QT}}(l, \bm x \smallsetminus x_{2l+1}) \widetilde
Z^{(2)}_{\mathrm{QT}}(l + 1; \bm x), \label{res1} \\[.5em]
& Z_{\mathrm{QT}}(4l - 1; \bm x) = [(-1)^{l+1} \sigma^{3l-2}(a)
\sigma^l(a^2)] Z^{(1)}_{\mathrm{QT}}(l, \bm x ) \widetilde
Z^{(2)}_{\mathrm{QT}}(l; \bm x \smallsetminus x_{2l}). \label{res2}
\end{align}
Using recursive relations (\ref{e:7}) and (\ref{2ndz}) and the
initial values
\[
Z^{(1)}_{\mathrm{QT}}(1; \bm x) = 1, \qquad \widetilde
Z^{(2)}_{\mathrm{QT}}(1; \bm x) = 1,
\]
it is not difficult to check that the right-hand sides of (\ref{res1}) and
(\ref{res2}) satisfy the initial condition~(\ref{ini}) and recursive
relations (\ref{2nd}) and (\ref{3rd}).

\section{\mathversion{bold}Special value of the parameter $a$ and
enumerations}

It turns out, that in the special case $a = \exp (\rmi \pi/3)$ one can
relate the functions $Z^{(1)}_{\mathrm{QT}}(1; \bm x)$ and $\widetilde
Z^{(2)}_{\mathrm{QT}}(1; \bm x)$ to the partition functions of the
square-ice models corresponding to all alternating-sign matrices  and to
the half-turn symmetric alternating-sign matrices of odd order.

If $a = \exp (\rmi\pi/3)$, then
\begin{equation}
\sigma(a^2 x) = -\sigma(\bar a x) = \sigma(a \bar x),
\label{iden}
\end{equation}
and recursive relations (\ref{e:7}) for $r=1$ becomes
\begin{equation}
\left. Z^{(1)}_{\mathrm{QT}}(l; \bm x) \right|_{x_i = a x_j} =
\prod_{\substack{k = 1 \\ k \ne i, j}}^{2l}
\sigma^2(a \bar x_j x_k) Z^{(1)}_{\mathrm{QT}}(l - 1; \bm x \smallsetminus
x_i \smallsetminus x_j).
\label{1stzspec}
\end{equation}

Recall that the partition function $Z(l; \bm x, \bm y)$ of the square-ice
model, corresponding to all alternating-sign matrices depends on $2l$
parameters $x_1, \ldots, x_l$ and $y_1, \ldots, y_l$ (see, for example
\cite{Kup02}). It is a centered Laurent polynomial of width $l-1$  in the
square of each of the variables $x_1, \ldots, x_l$ and $y_1, \ldots, y_l$,
satisfying the recursive relations
\begin{equation}
\label{4th}
\left. Z(l; \bm x, \bm y) \right|_{y_i = a x_j} = \sigma(a^2)
\prod_{\substack{k = 1 \\ k \ne j}}^l \sigma (a \bar x_j y_k)
\prod_{\substack{k = 1 \\ k \ne i}}^l \sigma (a^2 \bar x_k x_j)
Z(l - 1; \bm x \smallsetminus x_j, \bm y \smallsetminus y_i),
\end{equation}
where $i, j = 1, \ldots, l$ and $i \ne j$, with the initial value $Z(1; \bm
x, \bm y) = \sigma (a^2)$. It was shown in paper \cite{my1} that in the
case $a = \exp (\rmi\pi/3)$ this function is symmetric in the union of the
variables $x_1, \ldots, x_l$ and $y_1, \ldots, y_l$ (see also \cite{Ok3}
and references therein). Introducing the symmetric notations
\[
\label{note}
x_i = x_i, \quad x_{i+l} = y_i, \quad i=1,2,...,l,
\]
and taking into account identity (\ref{iden}) we write the recursive
relations for $Z(l, \bm x)$ as
\begin{equation}
\label{4thspec}
\left. Z(l; \bm x) \right|_{x_i = a x_j} = \sigma(a^2)
\prod_{\substack{k = 1 \\ k \ne i,j}}^{2l} \sigma (a \bar x_j x_k)
Z(l - 1; \bm x \smallsetminus x_i \smallsetminus x_j).
\end{equation}
Comparing relations (\ref{1stzspec}) with relations (\ref{4thspec}) and
taking into account the initial values for $Z^{(1)}_{\mathrm{QT}}(l; \bm
x)$ and $Z^2(l; \bm x)$, we find that
\begin{equation}
\label{result1}
Z^{(1)}_{\mathrm{QT}}(l; \bm x) = \sigma ^{-2l}(a^2) Z^2(l; \bm x).
\end{equation}
This equality was also obtained by Okada \cite{Ok3}.

The function $Z^{(2)}_{\mathrm{QT}}(l; \bm x)$ in the case $a=\exp
(\rmi\pi/3)$ satisfies the recursive relations
\begin{equation}
\left. \widetilde Z^{(2)}_{\mathrm{QT}}(l; \bm x) \right|_{x_i = a x_j} =
- \prod_{\substack{k = 1 \\ k \ne i, j}}^{2l-1} \sigma^2(a \bar x_j x_k)
\widetilde Z^{(2)}_{\mathrm{QT}}(l - 1; \bm x \smallsetminus x_i
\smallsetminus x_j),
\label{2ndzspec}
\end{equation}
which are rather different from (\ref{1stzspec}).

\begin{figure}[ht]
\centering
\begin{pspicture}(-1,-1)(10,12)
  \arrowLine(2,2)(2,0)
  \arrowLine(4,2)(4,0)
  \arrowLine(2,10)(2,12)
  \arrowLine(4,10)(4,12)
  \arrowLine(0,2)(2,2)
  \arrowLine(0,4)(2,4)
  \arrowLine(0,8)(2,8)
  \arrowLine(0,10)(2,10)
  \psline(2,2)(6,2)
  \psline(2,4)(6,4)
  \psline(2,8)(6,8)
  \psline(2,10)(6,10)
  \psline(2,2)(2,10)
  \psline(4,2)(4,10)
  \psarc(6,6){2}{270}{90}
  \psarc(6,6){4}{270}{90}
  \psdots(2,2)(2,4)(2,8)(2,10)(4,2)(4,4)
  \psdots(4,8)(4,10)(2,6)(4,6)(6,4)(6,2)
  \arrowLine(0,6)(2,6)
  \psline(2,6)(4,6)
  \psarc(4,4){2}{0}{90}
  \psline(6,4)(6,2)
  \arrowLine(6,2)(6,0)
  \rput(-0.5,10){$x_1$}
  \rput(-0.5,8){$x_2$}
  \rput(-0.5,6){$x_3$}
  \rput(-0.5,4){$x_2$}
  \rput(-0.5,2){$x_1$}
  \rput(2,-0.5){$y_1$}
  \rput(4,-0.5){$y_2$}
  \rput(6,-0.5){$y_3$}
\end{pspicture}
\caption{Square-ice corresponding to the half-turn  symmetric
alternating-sign matrices of odd order} \label{f:13}
\end{figure}
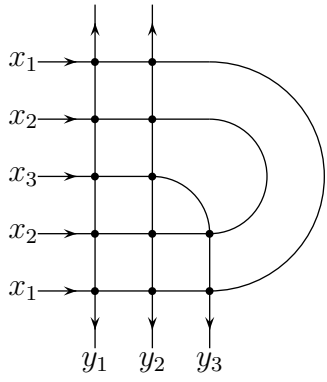

In our recent paper \cite{RazStr05} we considered the partition function
for the square-ice model corresponding to the half-turn symmetric
alternating-sign matrices of odd order (see Figure~\ref{f:13}).
The corresponding partition function $Z_{\mathrm{HT}}(2l-1; \bm x, \bm y)$
depends here on $2l$ spectral parameters $x_1, \ldots, x_l$ and $y_1,
\ldots, y_l$. We proved that in the case $a = \exp(\rmi\pi/3)$ and $y_{m+1}
= x_{m+1}$ the partiton function is a symmetric function in all $2l-1$
variables. Again introducing symmetric notations
\[
x_i = x_i, \quad i = 1, \ldots, l, \qquad x_{i+l} = y_i, \quad i = 1,
\ldots, l-1,
\]
and using for the function under consideration the same notation
$Z_{\mathrm{HT}}(2l - 1; \bm x)$, one sees that this function satisfies the
recursive relations
\begin{equation}
\label{5thspec}
\left. Z_{\mathrm{HT}}(2l - 1; \bm x) \right|_{x_i = a x_j}= \sigma^2(a^2)
\prod_{\substack{k=1 \\ k \ne i, j}}^{2l-1} \sigma^2(a \bar x_j x_k)
Z_{\mathrm{HT}}(2l - 3; \bm x \smallsetminus x_i \smallsetminus x_j),
\end{equation}
where $i, j = 1, \ldots, 2l-1$ and $i \ne j$.
Comparing relations (\ref{2ndzspec}) with relations (\ref{5thspec}) and
taking into account the initial values for $Z^{(2)}_{\mathrm{QT}}(l; \bm
x)$ and $Z_{\mathrm{HT}}(2l - 1; \bm x)$, we see that
\begin{equation}
\widetilde Z^{(2)}_{\mathrm{QT}}(l; \bm x) = (-1)^{l+1} \sigma^{2-2l}(a^2)
Z_{\mathrm{HT}}(2l - 1; \bm x).
\label{result2}
\end{equation}
Relations (\ref{result1}) and (\ref{result2}) allow us to write equalities
(\ref{res1}) and (\ref{res2}) as
\begin{align*}
\label{Zoddresult}
&Z_{\mathrm{QT}}(4l + 1; \bm x)= Z^2(l; \bm x \smallsetminus
x_{2l+1}) Z_{\mathrm{HT}}(2l + 1; \bm x), \\
&Z_{\mathrm{QT}}(4l - 1; \bm x)= Z^2(l; \bm x) Z_{\mathrm{HT}}(2l - 1; \bm
x \smallsetminus x_{2l}).
\end{align*}
It is natural to recall here the similar equality
\[
Z_{\mathrm{QT}}(4l; \bm x) = Z^2(l; \bm x) Z_{\mathrm{HT}}(2l; \bm x),
\]
obtained by Okada \cite{Ok3}.

Considering the last equalities at $\bm x = (1, \ldots, 1)$, one comes to
the relations
\[
A_{\mathrm{QT}}(4l + 1) =  A^2(l) A_{\mathrm{HT}}(2l + 1), \qquad
A_{\mathrm{QT}}(4l - 1) = A^2(l) A_{\mathrm{HT}}(2l - 1),
\]
where $A$ instead of $Z$ means the number of alternating-sign matrices
of the corresponding kind.
Combining these relations with the result obtained by Kuperberg for the
matrices of even order, we have
\[
A_{\mathrm{QT}}(4l + \epsilon)= A^2(l) A_{\mathrm{HT}}(2l + \epsilon),
\qquad \epsilon = -1, 0, 1.
\]
Thus, the Robbins conjecture \cite{Rob00} on the enumeration of the
quarter-turn symmetric alternating-sign matrices is proved.

{\it Acknowledgments}
The work was supported in part by the Russian Foundation for Basic Research
under grant \# 04--01--00352.

\end{document}